\documentclass[letters,usenatbib]{mnras}

\usepackage{newtxtext,newtxmath}
\usepackage[T1]{fontenc}
\usepackage{adjustbox}
\usepackage{subfigure}
\usepackage{hyperref}
\DeclareRobustCommand{\VAN}[3]{#2}
\let\VANthebibliography\thebibliography
\def\thebibliography{\DeclareRobustCommand{\VAN}[3]{##3}\VANthebibliography}

\def\be{\begin{equation}}
\def\ee{\end{equation}}
\def\ba{\begin{eqnarray}}
\def\ea{\end{eqnarray}}

\def\msun{M_\odot}
\def\ltsima{$\; \buildrel < \over \sim \;$}
\def\simlt{\lower.5ex\hbox{\ltsima}}
\def\gtsima{$\; \buildrel > \over \sim \;$}
\def\simgt{\lower.5ex\hbox{\gtsima}}

\usepackage{graphicx}	% Including figure files
\usepackage{amsmath}	% Advanced maths commands 
\usepackage{xcolor} 

\usepackage{rotating}
\usepackage{threeparttable}
\usepackage{float}
\definecolor{falured}{rgb}{0.5, 0.09, 0.09}

\NewDocumentCommand{\codeword}{v}{%
\texttt{\textcolor{black}{#1}}%
}
\title[Multi-temperature dust in Holmberg II]{Unravelling multi-temperature dust populations in the dwarf galaxy Holmberg II}

\author[Bordoloi et al.]{
Olag Pratim Bordoloi,$^{1}$\thanks{\href{mailto:olagpratim@gmail.com}{olagpratim@gmail.com}}
Yuri A. Shchekinov,$^{2}$\thanks{\href{mailto:yuri.and.s@gmail.com}{yuri.and.s@gmail.com}}
P. Shalima,$^{3}$\thanks{\href{mailto:shalima.p@gmail.com}{shalima.p@gmail.com}}
M. Safonova$^{4}$\thanks{\href{mailto:margarita.safonova62@gmail.com}{margarita.safonova62@gmail.com}}
and Rupjyoti Gogoi$^{1}$\thanks{\href{mailto:rupjyotigogoi@gmail.com}{rupjyotigogoi@gmail.com}}\\
$^{1}$Tezpur University, Napaam, Assam, India, 784028\\
$^{2}$Raman Research Institute, Bengaluru, India, 560080\\
$^{3}$Manipal Centre for Natural Sciences, Centre of Excellence, Manipal Academy of Higher Education, Manipal, Karnataka, India, 576104\\
$^{4}$Indian Institute of Astrophysics, Bengaluru, India, 560034\\
}

\date{Accepted XXX. Received YYY; in original form ZZZ}

\pubyear{2023}

\begin{document}
\label{firstpage}
\pagerange{\pageref{firstpage}--\pageref{lastpage}}
\maketitle

% Abstract of the paper
\begin{abstract}
Holmberg II – a dwarf galaxy in the nearby M81 group – is a very informative source of distribution of gas and dust in the
interstellar discs. High-resolution observations in the infrared (IR) allows us to distinguish isolated star-forming regions, photodissociation
(PDR) and HII regions, remnants of supernovae (SNe) explosions and, as such, can provide information about more relevant
physical processes. In this paper we analyse dust emission in the wavelength range  4.5 to  160 $\mu$m using the data from IR
space observatories at 27 different locations across the galaxy. We observe that the derived spectra can
be represented by multiple dust populations with different temperatures, which are found to be independent of their locations in the galaxy. By comparing the dust temperatures with the far ultraviolet (FUV) intensities observed by the UVIT instrument onboard AstroSat, we find that for locations showing a 100 $\mu$m peak, the temperature of cold (20 to 30 K) dust grains show a dependence on the FUV intensities, while such
a dependence is not observed for the other locations. We believe that the approach described here can be a good tool in revealing
different dust populations in other nearby galaxies with available high spatial resolution data.

\end{abstract}
 
 
%%%%%%%%%%%%%%%%% BODY OF PAPER %%%%%%%%%%%%%%%%%%

\section{Introduction}
 
Dwarfs galaxies (DG) -- galaxies with baryonic mass of $M_{b}$$\sim$$10^{9} M_{\odot}$, are commonly thought to be the building blocks for the entire hierarchy of mass distribution in the Universe, being progressed during its evolution since post-recombination epochs through merging process \citep{White1978}. Within this concept, DG in local Universe can serve as laboratories for understanding the details of physical processes regulating formation of the very first galaxies in early Universe \citep[see review by][and more recent discussions in \citet{Henkel2022,Annibali2022}]{Tolstoy2009}. Among the most important traits of DGs are their metal-poor interstellar medium (ISM) and, correspondingly, a low dust content \citep{Henkel2022} -- the features that are also commonly expected for early Universe galaxies. Observations of bright galaxies at $z>10$ conducted first by the {\it Hubble Space Telescope} (HST), and subsequently by the {\it James Webb Space Telescope} (JWST), have indeed indicated a deficient amount of dust in a set of high-$z$ galaxies \citep[see discussion in ][]{Finkelstein2022,Ferrara2023}.   

In order for local DGs to indeed provide relevant information for understanding of their more distant congeners, the comparative analysis of their global emission characteristics, along with the respective properties on smaller scales, is of a high importance \citep[relevant discussion can be found in][]{Izotov2021,Henkel2022}. A good example of a DG in local Universe is the Holmberg~II (Ho~II) galaxy belonging to the nearby M81 group at a distance of $\simeq 3.4$ Mpc. Its modest inclination angle ($\approx$$27^\circ$; \citet{Sanchez2014}) makes Ho~II a very informative source of distribution of gas and dust in the interstellar discs. Spitzer and Herschel space telescopes with the angular resolution of $\sim 2^{\prime\prime}-5''$ can probe the Ho~II interstellar disc with a high scrutiny. Scanning the interstellar medium (ISM) with such resolution can allow to distinguish isolated star forming regions (SFR), photodissociation (PDR) and HII regions, supernovae remnants, and can provide information about relevant physical processes. In this {\it Letter} we analyze spectra of dust emission in the range $\lambda\lambda=4.5, \,\ldots 160~\mu$m from Spitzer and Herschel archival data, focusing on a few rather small-area locations. It allows us to distinguish IR emission which cannot be attributed to the isothermal dust, thus requiring existence of at least two populations of dust with different temperatures.

Since the primary source of dust heating is the absorption of UV/optical photons by the dust grains, we have also utilised the highest available resolution ($1.2''-1.6''$) FUV observations of Ho~II galaxy obtained by the UVIT instrument of India's AstroSat mission \citep{Singh2014} as part of this study. The data correspond to 3 epochs in 2016, 2 epochs in 2019, and 3 epochs in 2020 \citep{Vinokurov2022}. These observations are crucial in understanding the UV radiation fields, responsible for modifying the dust populations and their thermal IR emission profiles at these locations. 
\vspace{-0.3in}

\section{Observational data}

We obtained the IR images of Ho~II from two surveys, namely the SINGS survey \citep{Kennicutt2003} and the KINGFISH survey \citep{Kennicutt2011}, archived at NASA/IPAC Infrared Science Archive, at 8 different wavelengths\footnote{To access SINGS and KINGFISH data, visit \url{https://irsa.ipac.caltech.edu/data/SPITZER/SINGS/galaxies/hoii/} and \url{https://irsa.ipac.caltech.edu/data/Herschel/KINGFISH/galaxies/HoII/PACS/}, respectively.}. 3.6 $\micron$, 4.5 $\micron$, 5.8 $\micron$, and 8 $\micron$ images were obtained by the IRAC (Infrared Array Camera) instrument of Spitzer Space Telescope\footnote{see \citet{Fazio2004} for details about IRAC instrument.}, while the MIPS (Multiband Imaging Photometer for Spitzer) instrument\footnote{see \citet{Rieke2004} for details about MIPS instrument.} provided 24 $\mu$m, 70 $\mu$m, and 160 $\mu$m images. The 100 $\micron$ image was obtained by the PACS (Photodetector Array Camera and Spectrometer) instrument onboard Herschel Space Observatory. A constant background level was already subtracted from each of the images.  

The images were further corrected to account for stellar contamination. For that we aligned and convolved the images to the same point-spread function (PSF) using IRAF\footnote{IRAF is distributed by the National Optical Astronomy Observatory, which is operated by the Association of Universities for Research in Astronomy, Inc., under a cooperative agreement with the National Science Foundation.} tasks \codeword{imalign} and \codeword{psfmatch}, respectively. The IR intensities were extracted from these aligned and convolved images at our selected locations by aperture photometry in $5''$-radius circular apertures. To account for the stellar contamination, we follow the methodology outlined by \citet{Helou2004}, by employing a scaling approach based on the assumption that the 3.6-$\micron$ emission represents the total stellar emission. We scaled down the 3.6 $\micron$ intensities by specific factors, i.e. 0.596, 0.399, 0.232 and 0.032 for 4.5 $\micron$, 5.8 $\micron$, 8 $\micron$, and 24 $\micron$,  respectively. These scaled 3.6 $\micron$ intensities were then subtracted from the respective intensities from the convolved images at corresponding wavelengths, which allowed us to isolate what we refer to as `diffuse dust emission'. 

No correction for stellar contamination was applied to 70 $\micron$, 100 $\micron$, and 160 $\micron$ images, as stellar contribution diminishes rapidly with increasing wavelength in the IR, and thus can be considered negligible in this range. Interestingly, we observed the lack of 8 $\micron$ emission in most regions, in line with the proposed deficiency of polycyclic aromatic hydrocarbons (PAH) in this galaxy \citep{Li2020}, except in three specific locations. Furthermore, the locations where IR emission was not detected in the range of wavelengths considered here were removed, resulting in a final sample of 27 locations. The IR intensities of this final set of locations, along with their respective errors, are presented in Table~\ref{tab:IR_intensities_part} and displayed on the FUV image taken on December 9, 2016 in Fig.~\ref{fig:locations_overplotted}. FUV intensities at these selected locations, extracted using the same 5$''$-radius aperture, showed to be approximately in the range of $(1.5-28)\times$10$^{-10}$ ergs cm$^{-2}$ s$^{-1}$ sr$^{-1}$ \AA$^{-1}$. A more detailed analysis of the FUV emission in Ho~II will be presented in the forthcoming paper \citep[in preparation]{Bordoloi2023}. 

\begin{table*}
\centering
\caption{Details of \textit{Spitzer} IR observations. This a sample of first few entries of the table; the full table is available online.}
\begin{adjustbox}{width=\textwidth, center}
    \begin{tabular}{cccccccccc}
    \hline
Location & \textit{l} & \textit{b}  & I$_{4.5 \micron}$ & I$_{5.8 \micron}$ & I$_{8 \micron}$ & I$_{24 \micron}$ & I$_{70 \micron}$ & I$_{100 \micron}$ & I$_{160 \micron}$\\
  No      & (degrees) & (degrees)  & (MJy sr$^{-1}$) & (MJy sr$^{-1}$) & (MJy sr$^{-1}$) & (MJy sr$^{-1}$) & (MJy sr$^{-1}$) & (MJy sr$^{-1}$) & (MJy sr$^{-1}$) \\
       \hline
\textcolor{red}{1} & \textcolor{red}{144.2731} & \textcolor{red}{32.6844} & 0.0106 $\pm$ 0.0273 & 0.0330 $\pm$ 0.0158 & 0 $\pm$ 0.0141 & 0.0333 $\pm$ 0.0100 & 1.9079 $\pm$ 0.0742 & 0.2300 $\pm$ 0.526 & 3.5553 $\pm$ 0.3761 \\
\textcolor{cyan}{2} & \textcolor{cyan}{144.2898} & \textcolor{cyan}{32.6750} & 0.0069 $\pm$ 0.0075 & 0.0200 $\pm$ 0.0079 & 0 $\pm$ 0.0062 & 0.1092 $\pm$ 0.0121 & 5.2336 $\pm$ 0.4443 & 7.2000 $\pm$ 0.5544 & 3.8332 $\pm$ 0.4575 \\
\textcolor{cyan}{3} & \textcolor{cyan}{144.2512} & \textcolor{cyan}{32.6607} & 0.0055 $\pm$ 0.0015 & 0.0122 $\pm$ 0.0035 & 0 $\pm$ 0.0065 & 0.6279 $\pm$ 0.1453 & 7.7358 $\pm$ 1.2877 & 14.4400 $\pm$ 2.5546 & 1.6263 $\pm$ 0.5952 \\
\textcolor{cyan}{4} &\textcolor{cyan}{144.2952} & \textcolor{cyan}{32.7203} & 0.0114 $\pm$ 0.0016 & 0.0191 $\pm$ 0.0035 & 0 $\pm$ 0.0043 & 0.3012 $\pm$ 0.0467 & 5.9099 $\pm$ 0.1316 & 8.7800 $\pm$ 0.8157 & 3.7606 $\pm$ 0.1606 \\
\textcolor{green}{5} &\textcolor{green}{144.2699} & \textcolor{green}{32.6985} & 0.0098 $\pm$ 0.0068 & 0.0340 $\pm$ 0.0064 & 0 $\pm$ 0.0053 & 0.1225 $\pm$ 0.0153 & 4.5948 $\pm$ 0.4802 & 1.9600 $\pm$ 0.4780 & 4.5118 $\pm$ 0.3977 \\
\textcolor{green}{6} &\textcolor{green}{144.2882} & \textcolor{green}{32.6942} & 0.0112 $\pm$ 0.0038 & 0.0245 $\pm$ 0.0037 & 0 $\pm$ 0.0045 & 0.0694 $\pm$ 0.0077 & 1.8620 $\pm$ 0.1317 & 2.4400 $\pm$ 0.4311 & 2.8699 $\pm$ 0.0896 \\
\textcolor{cyan}{7} &\textcolor{cyan}{144.2960}& \textcolor{cyan}{32.6858} & 0.0053 $\pm$ 0.0023 & 0.0171 $\pm$ 0.0040 & 0 $\pm$ 0.0041 & 0.0376 $\pm$ 0.0119 & 2.3811 $\pm$ 0.1101 & 3.8800 $\pm$ 0.3801 & 3.0756 $\pm$ 0.2645 \\
\textcolor{cyan}{8} &\textcolor{cyan}{144.2816} & \textcolor{cyan}{32.6822} & 0.0183 $\pm$ 0.0218 & 0.0619 $\pm$ 0.0147 & 0.0393 $\pm$ 0.0217 & 0.3235 $\pm$ 0.0269 & 5.0719 $\pm$ 0.3350 & 10.6100 $\pm$ 0.4589 & 5.0871 $\pm$ 0.0997 \\
\hline
\label{tab:IR_intensities_part}
\end{tabular}
\end{adjustbox}
\end{table*} 

\section{Results and Analysis} \label{sec:res}

\begin{figure*}
    \includegraphics[width=2.5in, height=2.5in]{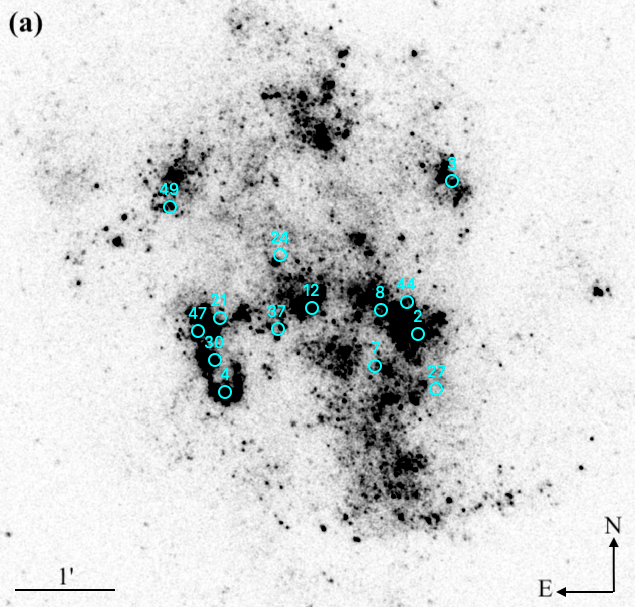}
    \includegraphics[width=2.5in, height=2.5in]{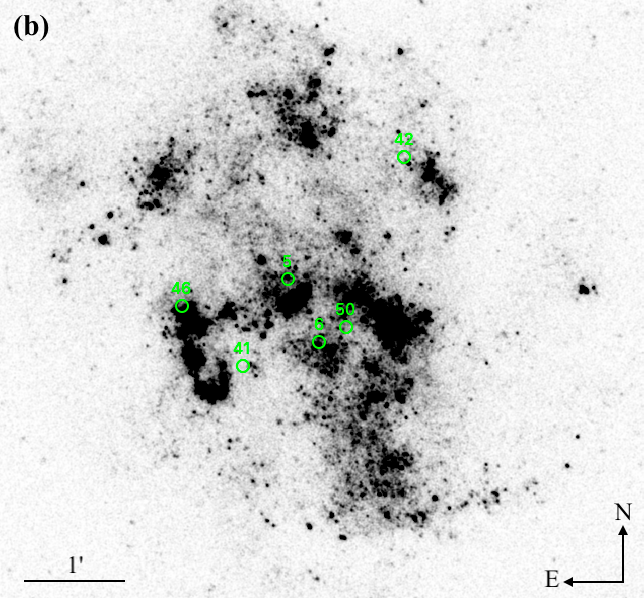}
    \includegraphics[width=2.5in, height=2.5in]{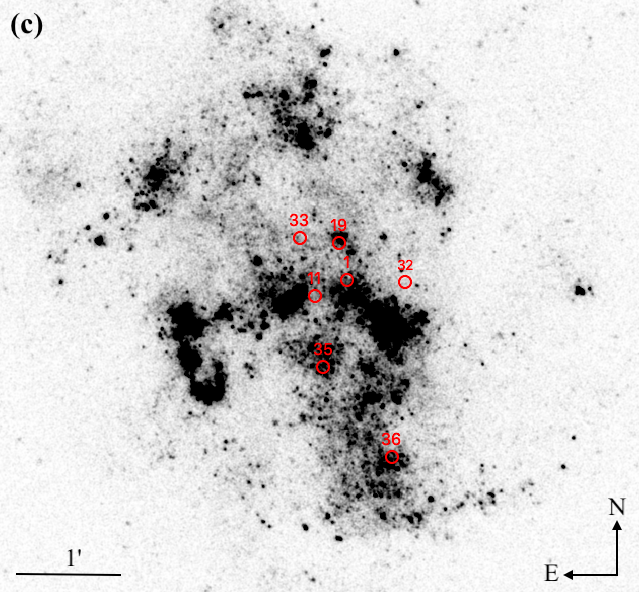}
    \caption{Locations depicted in Table~\ref{tab:IR_intensities_part} overplotted on the FUV image of Ho~II obtained by the UVIT instrument \citep{Vinokurov2022}. (a) Locations where intensity peaks at 100 $\micron$, represented by cyan colour, mostly lie in the SFR, (b) Locations where intensity peaks at 70 $\micron$, represented by green colour, lie mostly in the outskirts of SFR, (c) Locations lying in HI voids (N(HI) < $1 \times 10^{21}$ cm$^{-2}$, \citet{Walter2007}), represented by red colour. The circles representing the locations are of the same size as circular apertures ($5''$ radius) used to extract the IR intensities from the locations. The FUV image was taken on December 9, 2016 with an exposure time of around 9 ks.}
    \label{fig:locations_overplotted}
\end{figure*}

Locations depicted in Table~\ref{tab:IR_intensities_part}, and displayed in Fig.~\ref{fig:locations_overplotted}, can be broadly classified into the following categories:
\begin{enumerate} 
\item Locations with a peak intensity around 100 $\micron$. The intensities of these locations after 100 $\micron$ show a decreasing trend towards longer wavelengths. 
\item  Locations with peak intensity around 70 $\micron$. The intensities, after peaking around 70 $\micron$, mostly decrease towards 100 $\micron$, and then again show an increasing trend towards 160 $\micron$.
\item Locations corresponding to HI voids. These voids are the regions where the neutral hydrogen column density N(HI) is less than $1 \times 10^{21}$ cm$^{-2}$ \citep{Walter2007}. The intensities of these locations typically peak at 70 $\micron$, and then experience a dip at $100~\mu$m. 
\item Locations with an additional emission at 8 $\micron$. The intensity of these locations peaks around 100 $\micron$ and decreases towards longer wavelengths, similar to the locations in category (i).
\end{enumerate} 

\vspace{-0.2in}

\subsection{Dust in Holmberg II}

A comprehensive study of dust properties in their interplay with other physical characteristics in Holmberg II galaxy among other M81 dwarf galaxies is given by \cite{Walter2007}. The IR brightness at 24, 70, and 160~$\mu$m has been found to correlate with HI distribution, though with rather patchy and sparse distribution of the dust emission, which is either assembled into great complexes, or because it is heated by localized sources, such as HII regions, SNe, or stellar wind shocks \citep[e.g., Fig.~2 in][see also discussion in \citet{Galliano2018}]{Walter2007}. Following \citet{Galliano2021} and \cite{Egorov2023}, the latter would result in a suppressed emission in PAH molecules. This is consistent with the conclusion by \cite{Walter2007} that PAH features at 8~$\mu$m and 11.3~$\mu$m in Ho II contribute only around $\sim 0.1$ as compared to the continuum -- nearly 20 times lower than in their template galaxy NGC 7552. It is worth noting in this respect that the gas scale height in Ho~II is around 200 pc \citep{Banerjee} and, accounting for the fact that the galaxy has a rather modest SFR of  $\Sigma_{\rm SF}\sim 0.016~\msun~{\rm yr}^{-1}~{\rm kpc}^{-2}$ \citep{Walter2007}, one can conclude that all mechanical energy from stellar activity is locked within the ISM disc without expulsion of it outside \citep{Biman}. This can result in an enhanced anticorrelation between the SFR and PAH abundance in dwarf galaxies with low gas scale heights \citep[for more discussion, see][]{Galliano2018,Galliano2021}. 

On the other hand, patchiness in spatial distribution of dust in Ho II can be connected with an exceptionally low amount of dust, $M_{d}/M_{\rm HI}\sim 3\times 10^{-4}$, which is $\sim$20 times lower than in spirals \citep[see Fig.~11a in][]{Walter2007}, being consistent with its low gas metallicity, $12+\log({\rm O/H})\simeq 7.68$ \citep[see panel {\it d)} on Fig.~8 in][]{Galliano2021}. With $M_{\rm HI}\sim 6\times 10^8~\msun$, this gives $M_d\sim 2\times 10^5~\msun$. 

\begin{figure}
\begin{center}
\includegraphics[width=0.4\textwidth]{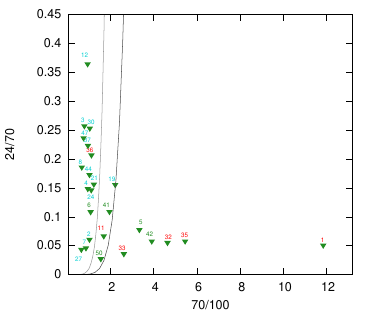}
\caption{$[\nu F_\nu]_{24~\mu{\rm m}}\big/[\nu F_\nu]_{70~\mu{\rm m}}$ vs. $[\nu F_\nu]_{70~\mu{\rm m}}\big/[\nu F_\nu]_{100~\mu{\rm m}}$ color--color diagram for the set of locations selected in this paper. Outliers towards high $[\nu F_\nu]_{70~\mu{\rm m}}\big/[\nu F_\nu]_{100~\mu{\rm m}}$ are apparently due to an enhanced ${70~\mu{\rm m}}$ emission. Thin grey line shows the ratio for $\beta=0$ (black body), thick darker line is for $\beta=2$, temperature changes from 3 K to 700 K from the bottom to top. Numbers and their colours correspond to locations shown in Fig.~\ref{fig:locations_overplotted}.}
\label{fig:color}
\end{center}
\end{figure}

More recently, \cite[][see their Fig.~2]{Zhou2016} payed attention to a principally different spectral shape of dust emission in metal deficient ($12+\log[{\rm O/H}]\leq 7.65$) dwarf galaxies, Ho~II among them, as compared to the galaxies with higher metallicity -- their colours, the ratios $[\nu F_\nu]_{70~\mu{\rm m}}\big/[\nu F_\nu]_{160~\mu{\rm m}}$ vs. $[\nu F_\nu]_{160~\mu{\rm m}}\big/[\nu F_\nu]_{250~\mu{\rm m}}$, look shifted towards lower values of the spectral index $\beta<1$. We found this tendency in the Ho~II galaxy on a colour--colour diagram $[\nu F_\nu]_{24~\mu{\rm m}}\big/[\nu F_\nu]_{70~\mu{\rm m}}$ vs. $[\nu F_\nu]_{70~\mu{\rm m}}/[\nu F_\nu]_{100~\mu{\rm m}}$, representing the dust at different locations in its disc (Fig.~\ref{fig:color}). It is clearly seen that some locations concentrate beyond the domain of positive spectral index $\beta\geq 0$. No less striking are also outliers with $[\nu F_\nu]_{24~\mu{\rm m}}\big/[\nu F_\nu]_{70~\mu{\rm m}}\!\sim 0.05$ and $[\nu F_\nu]_{70~\mu{\rm m}}\big/[\nu F_\nu]_{100~\mu{\rm m}}$, being as high as $\sim$10. In this regard, it is remarkable that these outliers come from the locations selected through different, though not mutually excluding, criteria. Locations marked in red represent regions deficient in atomic hydrogen ($N({\rm HI})<10^{21}$ cm$^{-2}$), in green -- those where IR spectra peak at 70 $\mu$m, and in cyan -- locations where spectra peak at 100 $\mu$m. \citet[][]{Draine2007} showed that stochastically heated small dust grains can contribute to an excessive 70 $\mu$m emission provided there is a sufficiently high UV energy density. However, as mentioned above, Ho~II is apparently deficient in small size grains and PAHs, and therefore in this case the emission could be due to stochastic heating of large dust grains by shock waves from stellar winds and SNe \citep[][see also latest examples in \citet{Drozdov2019,Drozdov2021}]{Dwek1986}. One therefore can assume that the outliers shown in Fig. \ref{fig:color} represent dust clumps heated by the shocks.   

\subsection{Multi-temperature dust spectra } 

In this {\it Letter} we interpret the differences observed in loosely selected IR sources in Sec. \ref{sec:res} as due to contributions from multi-temperature dust components that originate in distinct regions within the same field of view (FOV) under the dust heating from different not always resolved sources. 

The observed flux density from a cloud containing dust of total mass $M_d$ at frequency $\nu$ is given by \citep[see][]{Hildebrand1983}, 
\be 
F_\nu=\frac{3M_d}{4\pi a \rho_d D^2}Q(a,\nu)B_\nu(T)\,, 
\ee 
where $a$ is the dust grain radius (assumed $a$$\sim$$0.1~\mu$m), $\rho_d$ is the density of the dust material, $B_\nu(T)$ the Planck function, and $D$ is the detector--cloud separation. $Q(a,\nu)\simeq 0.01a({\nu}/{c})^{\alpha}$ is the emissivity where $a$ and $\nu$ are in cgs units. \cite{Dwek1992} assume a spectral index $\alpha=1.94$, though a wider range of $\alpha=1.5-2$ is commonly applied \citep[e.g.][]{Draine1984,Cortese2012}. For the sake of simplification of the fitting procedure and accounting of observational uncertainties, we assume $\alpha=2$, unless specified.

\begin{figure*}
    \centering
    \includegraphics[scale=0.40]{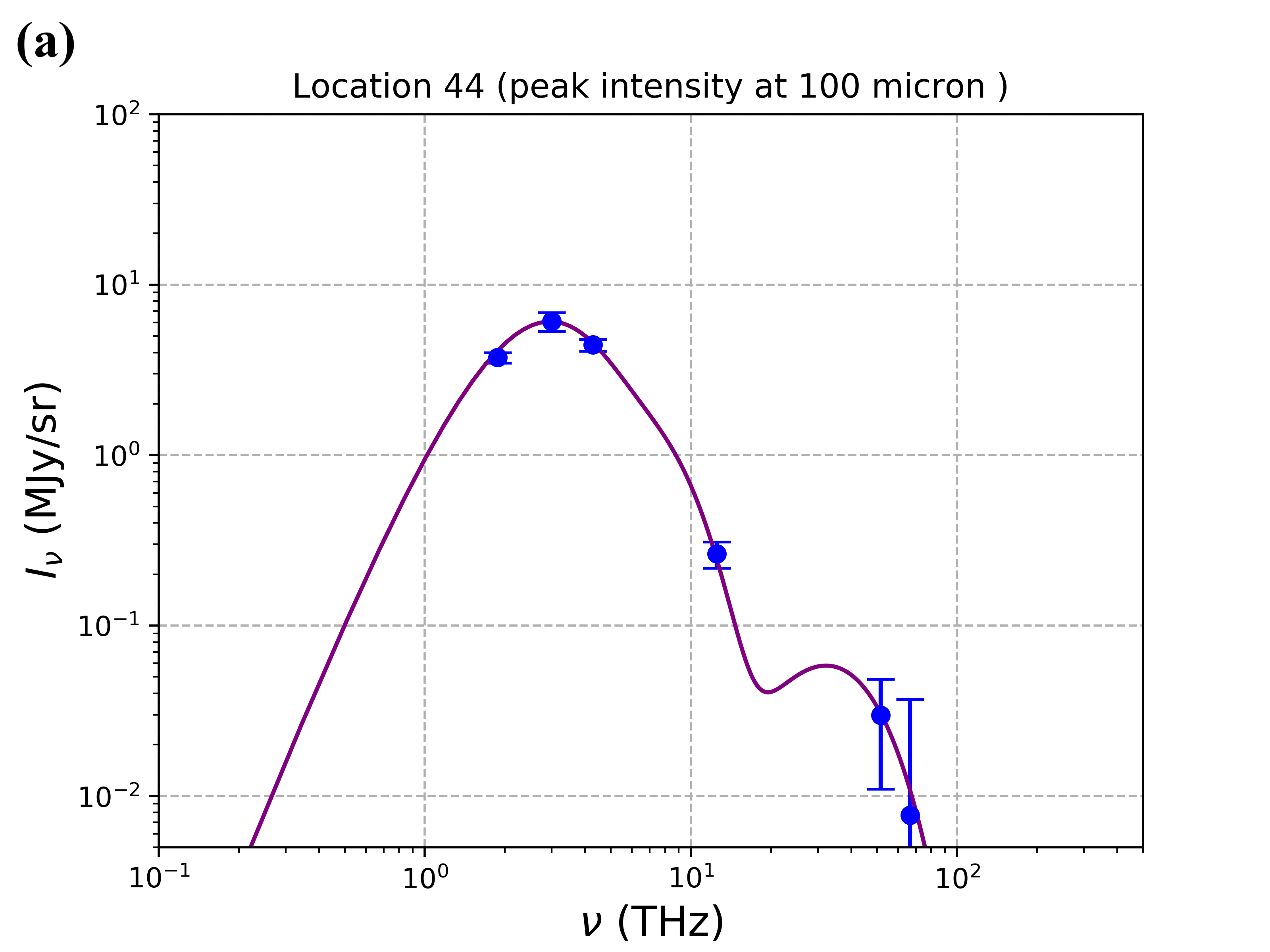}
    \includegraphics[scale=0.40]{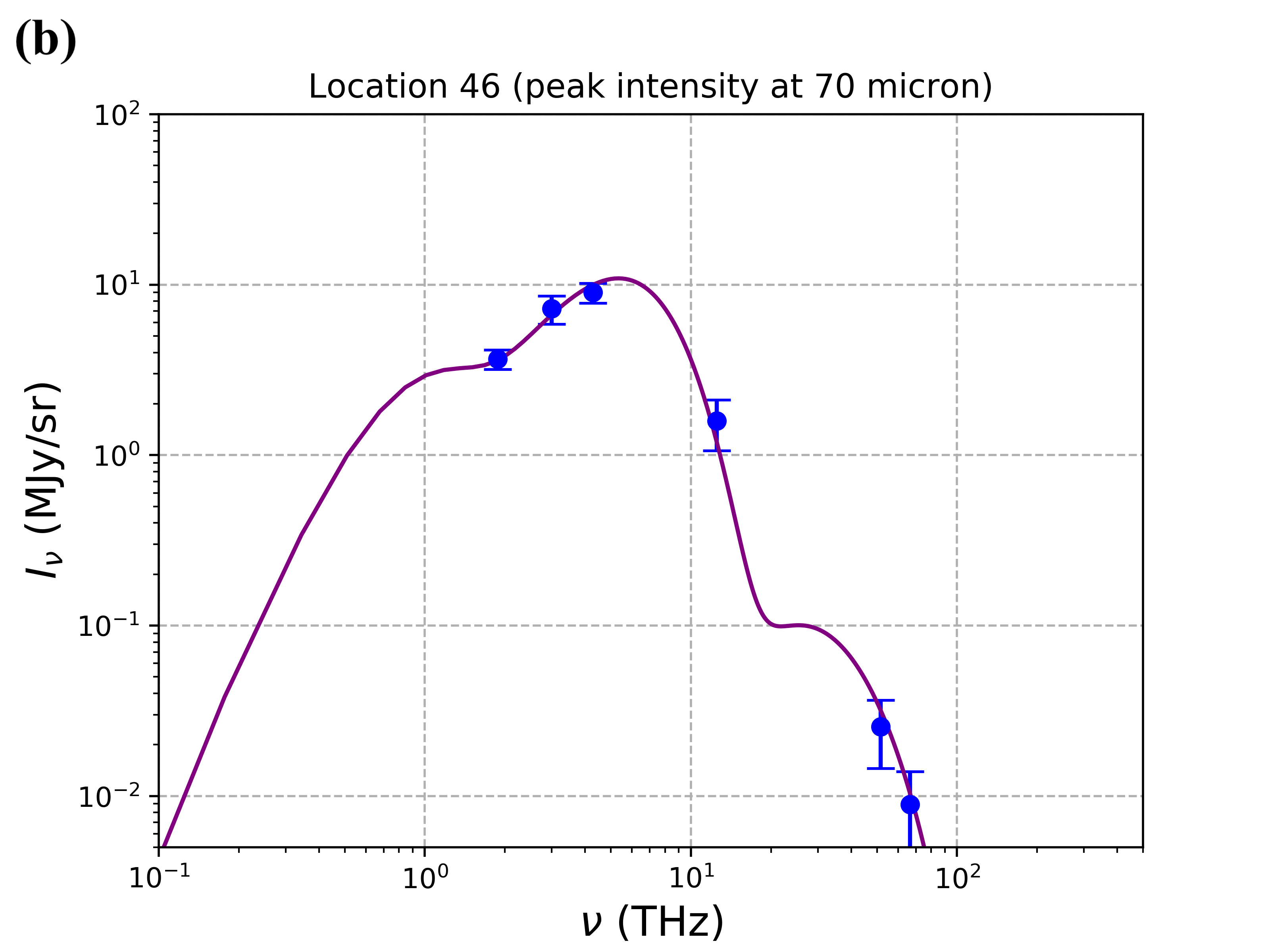}
    \includegraphics[scale=0.40]{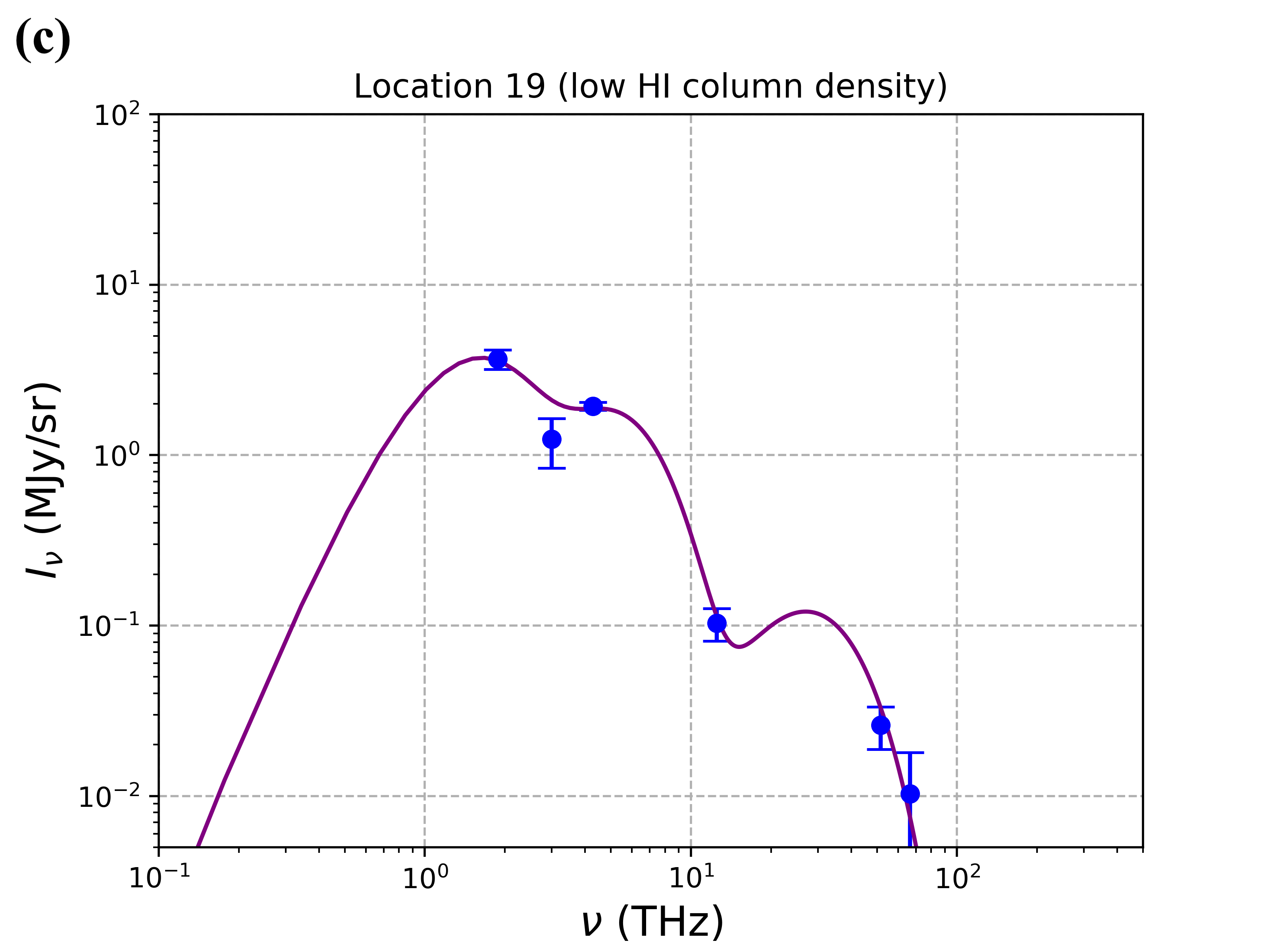}
    \includegraphics[scale=0.40]{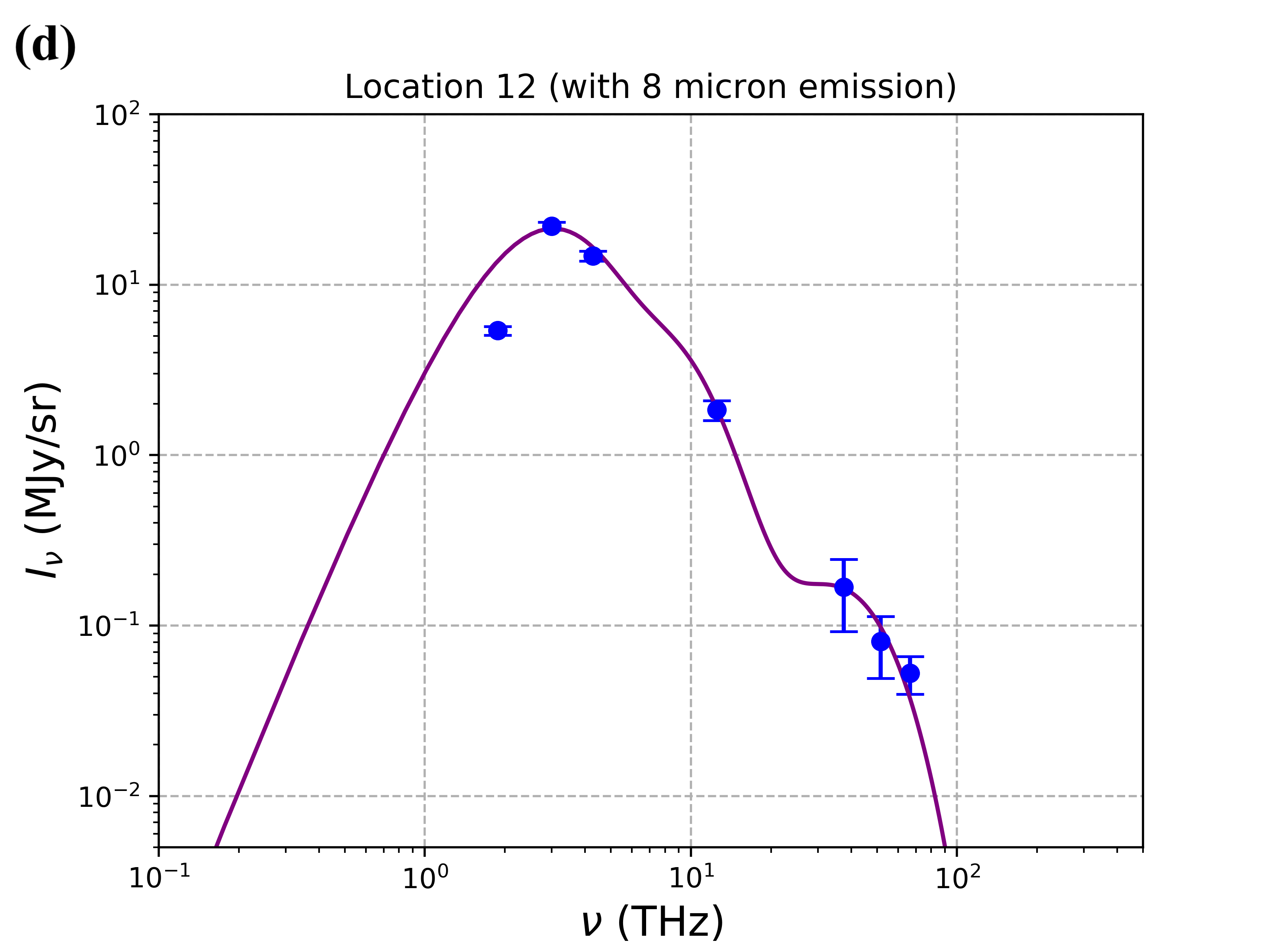}
\caption{Fits of the spectra of dust at different environmental conditions as classified in Fig.~\ref{fig:locations_overplotted}. They represent spectra of dust: {\bf i)} spectra peaking at 100 $\micron$,  {\bf ii)} spectra peaking at 70 $\micron$, {\bf iii)} in regions with HI column densities below the detection limit $N({\rm HI})<10^{21}$ cm$^{-2}$, {\bf iv)} spectra with emission at 8 $\micron$, {\bf v}) spectra with a weak intensity dip at 100 $\micron$ and {\bf vi}) spectra with a strong intensity dip at 100 $\micron$. Specifically, the presented spectra fitted with multi-temperature modified black body with the following temperatures and dust surface mass densities for cold components and dust masses for warm and hot ones: (a) $T = 27, 56, 310$ K ($\Sigma_d=6\times 10^{-5}~\msun$pc$^{-2}$, $M_d=0.009,~6.4\times 10^{-8}\msun$); (b) $T = 10, 25, 53, 220, 320$ K ($\Sigma_d=3\times 10^{-3}\msun$pc$^{-2}$, $M_d=0.5,~0.08,~4\times 10^{-7},~4\times 10^{-8}~\msun$); (c) $T = 15, 46, 260$ K ($\Sigma_d=7\times 10^{-4}~\msun$pc$^{-2}$, $M_d=0.025,~3\times 10^{-7}~\msun$); (d) $T = 28, 63, 100, 320$ K ($\Sigma_d=6\times 10^{-4}~\msun$pc$^{-2}$, $M_d=63,~0.002,~2\times 10^{-7}~\msun$). }

\label{fig:4locations}
\end{figure*}

Fig.~\ref{fig:4locations} illustrates several examples of the IR spectra observed in 4 specific locations of Ho~II galaxy (these locations are marked in Fig.~\ref{fig:locations_overplotted}). The fitting procedure is applied to the spectra in a given FOV within the error bars, with free variables being the dust temperature $T_d$ and dust mass $M_d$ under fixed spectral index $\beta=2$. The dust mass $M_d$ is reduced to the surface dust mass within the corresponding aperture with solid angle $\Delta\Omega$. Therefore, the fitting outcome is the dust temperature $T_d$, the dust mass $M_d$ for warm and hot ($T_d>30$ K) dust components, being presumably compact dusty structures, and the surface dust mass $\Sigma_d$ for colder ($T_d<30$ K) dust redistributed most likely in diffuse ISM.

From the estimates one can preliminary conclude that {\bf\it i)} cold dust at $T_d\simeq 10-30$ K is from the diffuse ISM with a dust-to-gas mass fraction, $\zeta_d=(2-3)\times 10^{-3}$ which is consistent with an order of magnitude lower metallicity, in estimates a gas scale height of $z_0=200$ pc and in the plane gas density of $n=1$ cm$^{-3}$ are assumed, {\bf\it ii)} dust at 30 to 50 K can apparently stem from the ISM in close vicinity of individual stars. The inferred surface dust mass density for the colder $T_d<30$ K components is on average $\langle\Sigma_d\rangle\sim (1-3)\times 10^{-3}~\msun$ pc$^{-2}$ and results in the total dust mass within 1.5$^{\prime}$ diameter of Ho II to be around $\sim$$(1-30)\times 10^4~\msun$, which is consistent with \citet{Walter2007}. The `high-temperature' peaks with $T_d > 220$ K require apparently alternative energy sources, such as the IR emission from dusty envelopes around red giants \citep{Jura1999,Lagadec2005},
circumstellar discs \citep{Uzpen2008}, as well as dusty clumps in wind-blown bubbles and SNe.  
Heating by stellar UV/FUV photons looks unphysical because the required energy density $\nu u_\nu$ is as high as $\sim 10^7$ of Habing value $\nu u_{\nu,0}=4\times 10^{-14}$ erg~cm$^{-3}$, the standard interstellar radiation field  \citep[Sec. 12.5 in ][]{Draine2011}.   

\section{Discussion and conclusions} 
\label{sec:dac}

\subsection{Discussion} 
 
Detailing of dust thermal emission on smaller spatial scales allows us to avoid blurring and confusing effects caused by averaging over the larger areas and mixing contributions from sources with different physical characteristics. Such a high resolution scrutiny is provided by the Spitzer and Herschel space telescopes in the nearby dwarf Ho II galaxy. Its relatively small inclination allows to recognize contributions from rather small scale features without considerable mixing that can originate along sight of lines in a crowd environment. It is worth noting that this approach, once completed, promises a good tool for revealing different dust populations in other nearby galaxies where high spatial resolution is reachable (e.g., M31, LMC, SMC, and other DGs in the Local Group).

In this {\it Letter}, amongst the IR spectra at locations of visually different nature as described above (see Fig.~\ref{fig:4locations} and a brief description in the caption), we recognized cloud(let)s with evidently different spectra, although not connected directly to differences  between the three groups of sources shown in Fig. \ref{fig:locations_overplotted}. For instance, IR  spectra in HI voids do not show obvious distinct features. In compact regions lying very close to the FUV bright spots, presumably associated with active star formation (although not always being tightly connected with bright H$\alpha$ regions resolved by the HST as, e.g., locations 8, 44 and 49), we have identified three populations of cloudlets with apparently cold, warm, and hot components. Within the simplest assumptions, we attributed these differences to different temperatures. Following \citet[][]{Walter2007}, we have neglected the contribution from PAH molecules. The primary physical reason for such an approximation is that PAH molecules are very fragile under strong FUV radiation \citep[see for recent discussions][and references therein]{Egorov2023}. 

\begin{figure}
\begin{center}
\includegraphics[scale=0.4]{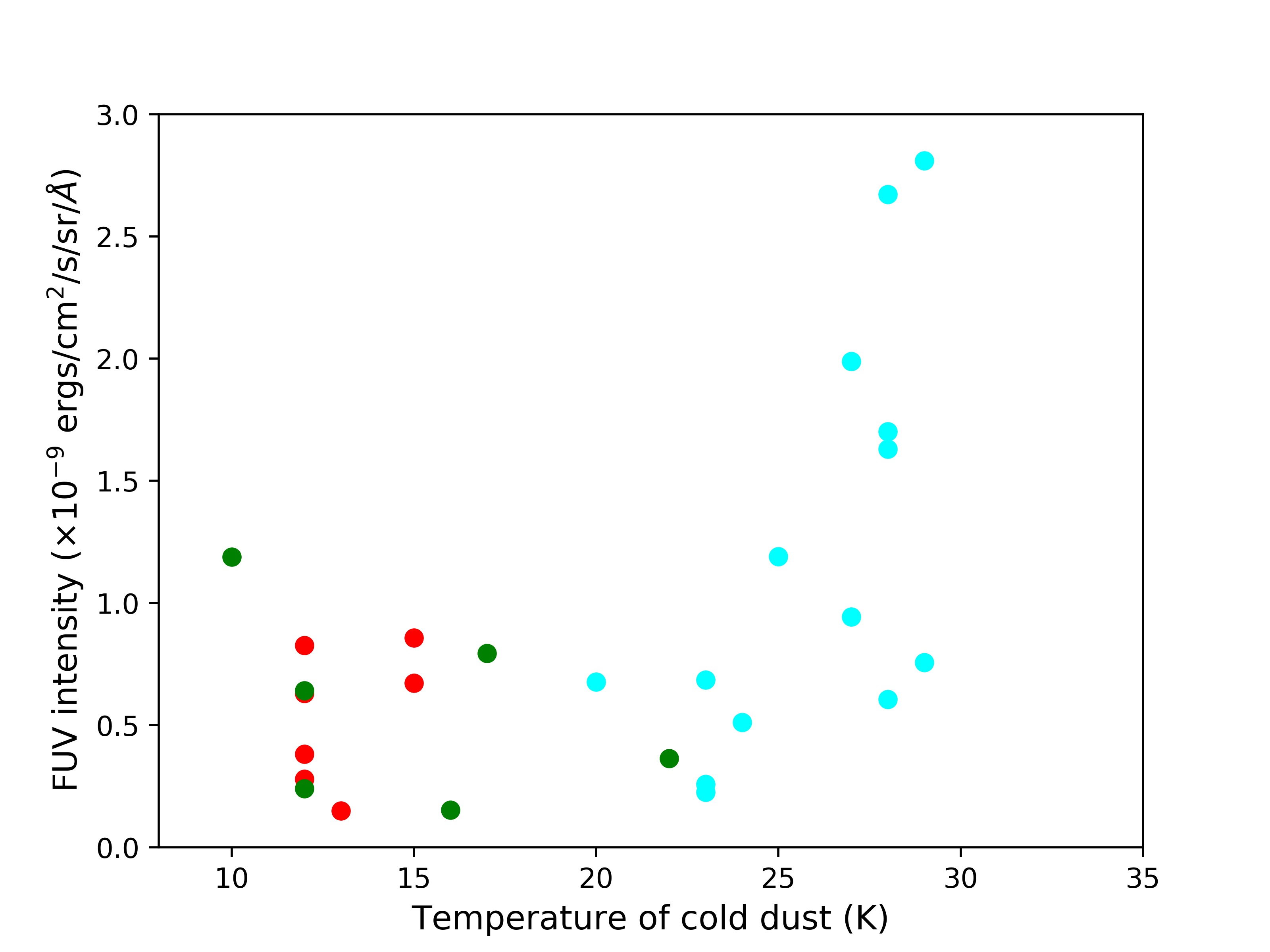}
\caption{Variation of temperature of cold dust with FUV intensity for the considered 27 locations. Colour coding is according to Fig.~\ref{fig:locations_overplotted}.}
\label{fig:fuvdust}
\end{center}
\end{figure}

It seems natural to assume that the coldest dust components are distributed diffusely throughout the ISM and are heated by the FUV radiation field. Surprisingly, preliminary inspection of possible connection of the cold dust population with the FUV radiation field has shown no obvious signs of correlation for most of the locations (Fig.~\ref{fig:fuvdust}). It is seen that temperatures of the cold dust components in HI voids, and in locations with spectra peaking at 70 $\mu$m, range between $10-20$ K and are not correlated with FUV intensity. However, the dust component of the sample with the peak intensity at 100 $\mu$m shows slightly higher temperatures $\approx 20-30$ K growing with the FUV intensity. 

It is pertinent to note that in the cleaned spectra with the flux at $8~\mu$m being clearly above the detection limit (locations 8, 12, 37 in Table~\ref{tab:IR_intensities_part}), we detect signs of enhanced intensity at wavelengths near those that are typically attributed to PAH features, i.e. $\sim\!3.3~\mu$m, $\sim\!6~\mu$m, and $\sim\!24~\mu$m. It is also worth noting that emission in this wavelength range is widely present in PDRs due to the main H$_2$ rotational lines at $\lambda\lambda=6.9,\,\ldots 17.0~\mu$m \citep{Compiegne2008}. However, in the case of Ho~II, we did not find any reference in the literature to these features being detected or studied. We are not aware of any reported data on these spectral features in this wavelength range for Ho~II galaxy since 2010. Therefore, within the above assumptions and following the arguments by \citet[][]{Walter2007,Egorov2023}, we approximated the spectra in this wavelength range by a modified black body from sources with higher temperatures. We attribute them to compact energy sources like dusty envelopes around red giant stars, including possibly, carbon stars, and dusty clumps heated by shock waves from stellar wind or SNe.

\subsection{Conclusions}

\begin{enumerate}
\item We present for the first time analysis of IR dust emission in the galaxy Holmberg~II to detect spatial variations of dust parameters on small physical scales of $\sim$ 82 pc (corresponding to a circular area of 5$''$-radius) using high angular resolution data from Spitzer and Herschel. 
\item In several locations, connected to physically distinct regions, we found spectra that can represent several -- up to five, dust populations with different temperatures. Spectral characteristics are not sensitive to the HI column density, except for the cold dust component with $T_d= 10-15$ K which concentrates predominantly in HI deficient regions. Preliminary inspection shows that this cold dust population does not show a dependence on FUV intensity.  
\item Similarly, the cold dust in those regions with spectra peaking at 70 $\mu$m has temperatures nearly in the same range as the HI voids with little dependence on FUV intensity. However, for those locations with the peak intensity at 100 $\mu$m, the temperature of the cold dust component ($T_d=20-30$ K) increases with FUV intensity. 
\item The estimated dust mass manifests signs of anti-correlation with its temperature, as formerly reported for dwarf star-forming galaxies by \cite{Izotov2014}. 
\end{enumerate}

\section*{Acknowledgements}

OPB and RG are thankful to Science \& Engineering Research Board (SERB), Department of Science \& Technology (DST), Government of India for financial support (EMR/2017/003092). OPB acknowledges the help received during this work from his colleagues Anshuman Borgohain and Hritwik Bora. SP acknowledges Manipal Centre for Natural Sciences, Centre of Excellence, Manipal Academy of Higher Education (MAHE) for facilities and support. MS acknowledges the financial support by the DST, Government of India, under the Women Scientist Scheme (PH) project reference number SR/WOS-A/PM-17/2019. YS acknowledges the hospitality of the Raman Research Institute, Bengaluru, India. 
 
\section*{Data Availability}

The data (complete Table~1 and an additional table on the insights to the selected locations) associated with this manuscript are available at the following website: \url{https://github.com/olagpratim/Details-of-Spitzer-IR-observations/blob/a9b835647bfbe02f85f2fd4e75abb6e145dd095b/AppendixA.pdf}

%%%%%%%%%%%%%%%%%%%% REFERENCES %%%%%%%%%%%%%%%%%% 

\bibliographystyle{mnras}
\bibliography{example}  

%%%%%%%%%%%%%%%%% APPENDICES %%%%%%%%%%%%%%%%%%%%%
\onecolumn
\appendix

\section{IR intensities and additional data for selected 27 locations}
%\twocolumn

\begin{table*}
\centering
\caption{IR intensities in selected 27 locations. Colour coding is according to Fig.~\ref{fig:locations_overplotted}. See Sec.~2 for details of calculations.}
\begin{adjustbox}{width=\textwidth, center}
\begin{tabular}{cccccccccc}
\hline
Location & \textit{l} & \textit{b}  & I$_{4.5 \micron}$ & I$_{5.8 \micron}$ & I$_{8 \micron}$ & I$_{24 \micron}$ & I$_{70 \micron}$ & I$_{100 \micron}$ & I$_{160 \micron}$\\
  No      & (degrees) & (degrees)  & (MJy sr$^{-1}$) & (MJy sr$^{-1}$) & (MJy sr$^{-1}$) & (MJy sr$^{-1}$) & (MJy sr$^{-1}$) & (MJy sr$^{-1}$) & (MJy sr$^{-1}$) \\
       \hline
\textcolor{red}{1} & \textcolor{red}{144.2731} & \textcolor{red}{32.6844} & 0.0106 $\pm$ 0.0273 & 0.0330 $\pm$ 0.0158 & 0 $\pm$ 0.0141 & 0.0333 $\pm$ 0.0100 & 1.9079 $\pm$ 0.0742 & 0.2300 $\pm$ 0.526 & 3.5553 $\pm$ 0.3761 \\
\textcolor{cyan}{2} & \textcolor{cyan}{144.2898} & \textcolor{cyan}{32.6750} & 0.0069 $\pm$ 0.0075 & 0.0200 $\pm$ 0.0079 & 0 $\pm$ 0.0062 & 0.1092 $\pm$ 0.0121 & 5.2336 $\pm$ 0.4443 & 7.2000 $\pm$ 0.5544 & 3.8332 $\pm$ 0.4575 \\
\textcolor{cyan}{3} & \textcolor{cyan}{144.2512} & \textcolor{cyan}{32.6607} & 0.0055 $\pm$ 0.0015 & 0.0122 $\pm$ 0.0035 & 0 $\pm$ 0.0065 & 0.6279 $\pm$ 0.1453 & 7.7358 $\pm$ 1.2877 & 14.4400 $\pm$ 2.5546 & 1.6263 $\pm$ 0.5952 \\
\textcolor{cyan}{4} &\textcolor{cyan}{144.2952} & \textcolor{cyan}{32.7203} & 0.0114 $\pm$ 0.0016 & 0.0191 $\pm$ 0.0035 & 0 $\pm$ 0.0043 & 0.3012 $\pm$ 0.0467 & 5.9099 $\pm$ 0.1316 & 8.7800 $\pm$ 0.8157 & 3.7606 $\pm$ 0.1606 \\
\textcolor{green}{5} &\textcolor{green}{144.2699} & \textcolor{green}{32.6985} & 0.0098 $\pm$ 0.0068 & 0.0340 $\pm$ 0.0064 & 0 $\pm$ 0.0053 & 0.1225 $\pm$ 0.0153 & 4.5948 $\pm$ 0.4802 & 1.9600 $\pm$ 0.4780 & 4.5118 $\pm$ 0.3977 \\
\textcolor{green}{6} &\textcolor{green}{144.2882} & \textcolor{green}{32.6942} & 0.0112 $\pm$ 0.0038 & 0.0245 $\pm$ 0.0037 & 0 $\pm$ 0.0045 & 0.0694 $\pm$ 0.0077 & 1.8620 $\pm$ 0.1317 & 2.4400 $\pm$ 0.4311 & 2.8699 $\pm$ 0.0896 \\
\textcolor{cyan}{7} &\textcolor{cyan}{144.2960}& \textcolor{cyan}{32.6858} & 0.0053 $\pm$ 0.0023 & 0.0171 $\pm$ 0.0040 & 0 $\pm$ 0.0041 & 0.0376 $\pm$ 0.0119 & 2.3811 $\pm$ 0.1101 & 3.8800 $\pm$ 0.3801 & 3.0756 $\pm$ 0.2645 \\
\textcolor{cyan}{8} &\textcolor{cyan}{144.2816} & \textcolor{cyan}{32.6822} & 0.0183 $\pm$ 0.0218 & 0.0619 $\pm$ 0.0147 & 0.0393 $\pm$ 0.0217 & 0.3235 $\pm$ 0.0269 & 5.0719 $\pm$ 0.3350 & 10.6100 $\pm$ 0.4589 & 5.0871 $\pm$ 0.0997 \\
\textcolor{red}{11} &\textcolor{red}{144.2757} & \textcolor{red}{32.6922} & 0.0094 $\pm$ 0.0045 & 0.0361 $\pm$ 0.0043 & 0 $\pm$ 0.0040 & 0.1135 $\pm$ 0.0067 & 4.9482 $\pm$ 0.3170 & 4.1600 $\pm$ 0.7900 & 1.8229 $\pm$ 0.7535 \\
\textcolor{cyan}{12} &\textcolor{cyan}{144.2774} &\textcolor{cyan}{32.6973} & 0.0527 $\pm$ 0.0132 & 0.0809 $\pm$ 0.0320 & 0.1682 $\pm$ 0.0759 & 1.8455 $\pm$ 0.2469 & 14.7611 $\pm$ 1.0000 & 22.1000 $\pm$ 1.2138 & 5.3792 $\pm$ 0.3122 \\
\textcolor{red}{19} &\textcolor{red}{144.2630} & \textcolor{red}{32.6846} & 0.0103 $\pm$ 0.0077 & 0.0261 $\pm$ 0.0073 & 0 $\pm$ 0.0054 & 0.1034 $\pm$ 0.0222 & 1.9352 $\pm$ 0.1041 & 1.2400 $\pm$ 0.4014 & 3.6667 $\pm$ 0.4822 \\
\textcolor{cyan}{21} &\textcolor{cyan}{144.2754} &\textcolor{cyan}{32.7183} & 0.0115 $\pm$ 0.0082 & 0.0327 $\pm$ 0.0056 & 0 $\pm$ 0.0059 & 0.5014 $\pm$ 0.1046 & 9.3478 $\pm$ 0.7582 & 10.7500 $\pm$ 0.3933 & 5.1692 $\pm$ 0.1822 \\
\textcolor{cyan}{24} &\textcolor{cyan}{144.2618} &\textcolor{cyan}{32.7021} & 0.0150 $\pm$ 0.0328 & 0.0467 $\pm$ 0.0212 & 0 $\pm$ 0.0142 & 0.1520 $\pm$ 0.0269 & 3.0249 $\pm$ 0.2186 & 3.8500 $\pm$ 0.6202 & 3.1424 $\pm$ 0.8492 \\
\textcolor{cyan}{27} &\textcolor{cyan}{144.3054} &\textcolor{cyan}{32.6733} & 0.0004 $\pm$ 0.0608 & 0.0216 $\pm$ 0.0361 & 0 $\pm$ 0.0205 & 0.0246 $\pm$ 0.0077 & 1.6464 $\pm$ 0.0832 & 3.5500 $\pm$ 0.4701 & 1.9316 $\pm$ 0.1883 \\
\textcolor{cyan}{30} &\textcolor{cyan}{144.2863} &\textcolor{cyan}{32.7213} & 0.0204 $\pm$ 0.0054 & 0.0278 $\pm$ 0.0070 & 0 $\pm$ 0.0089 & 0.7002 $\pm$ 0.0429 & 8.0550 $\pm$ 0.3701 & 10.9100 $\pm$ 0.4139 & 2.9251 $\pm$ 0.2354 \\
\textcolor{red}{32} &\textcolor{red}{144.2768} & \textcolor{red}{32.6716} & 0.0075 $\pm$ 0.0048 & 0.0249 $\pm$ 0.0049 & 0 $\pm$ 0.0048 & 0.0328 $\pm$ 0.0076 & 1.7242 $\pm$ 0.2066 & 0.5300 $\pm$ 0.4317 & 2.0164 $\pm$ 0.3426 \\
\textcolor{red}{33} &\textcolor{red}{144.2595} & \textcolor{red}{32.6931} & 0.0089 $\pm$ 0.0061 & 0.0408 $\pm$ 0.0052 & 0 $\pm$ 0.0058 & 0.0184 $\pm$ 0.0074 & 1.4919 $\pm$ 0.1011 & 0.8100 $\pm$ 0.3886 & 1.9019 $\pm$ 0.1035 \\
\textcolor{red}{35} &\textcolor{red}{144.2949} & \textcolor{red}{32.6935} & 0.0074 $\pm$ 0.0039 & 0.0372 $\pm$ 0.0062 & 0 $\pm$ 0.0045 & 0.0258 $\pm$ 0.0092 & 1.2994 $\pm$ 0.2115 & 0.3400 $\pm$ 0.4879 & 2.8701 $\pm$ 0.4514 \\
\textcolor{red}{36} &\textcolor{red}{144.3222} & \textcolor{red}{32.6821} & 0.0053 $\pm$ 0.0047 & 0.0048 $\pm$ 0.0046 & 0 $\pm$ 0.0040 & 0.0269 $\pm$ 0.0070 & 0.3790 $\pm$ 0.1105 & 0.4800 $\pm$ 0.3691 & 0.9162 $\pm$ 0.2116 \\
\textcolor{cyan}{37} &\textcolor{cyan}{144.2812} &\textcolor{cyan}{32.7059} & 0.0283 $\pm$ 0.0248 & 0.0783 $\pm$ 0.0157 & 0.0355 $\pm$ 0.0161 & 0.4197 $\pm$ 0.0166 & 4.7531 $\pm$ 0.1707 & 8.4700 $\pm$ 0.6406 & 5.6591 $\pm$ 0.2748 \\
\textcolor{green}{41} &\textcolor{green}{144.2905} & \textcolor{green}{32.7123} & 0.0065 $\pm$ 0.0032 & 0.0361 $\pm$ 0.0038 & 0 $\pm$ 0.0043 & 0.1141 $\pm$ 0.0109 & 3.0536 $\pm$ 0.3157 & 2.2200 $\pm$ 0.3941 & 2.5669 $\pm$ 0.1019 \\
\textcolor{green}{42} &\textcolor{green}{144.2438} &\textcolor{green}{32.6674} & 0.0029 $\pm$ 0.0022 & 0.0021 $\pm$ 0.0040 & 0 $\pm$ 0.0049 & 0.0256 $\pm$ 0.0064 & 1.2904 $\pm$ 0.1688 & 0.4700 $\pm$ 0.3980 & 1.2432 $\pm$ 0.3501 \\
\textcolor{cyan}{44} &\textcolor{cyan}{144.2809} &\textcolor{cyan}{32.6760} & 0.0077 $\pm$ 0.0293 & 0.0298 $\pm$ 0.0188 & 0 $\pm$ 0.0131 & 0.2634 $\pm$ 0.0461 & 4.4406 $\pm$ 0.3565 & 6.1000 $\pm$ 0.7717 & 3.7378 $\pm$ 0.2591 \\
\textcolor{green}{46} &\textcolor{green}{144.2714} &\textcolor{green}{32.7231} & 0.0089 $\pm$ 0.0050 & 0.0255 $\pm$ 0.0110 & 0 $\pm$ 0.0166 & 1.5858 $\pm$ 0.5204 & 8.9976 $\pm$ 1.1983 & 7.2400 $\pm$ 1.3546 & 3.6638 $\pm$ 0.4835 \\
\textcolor{cyan}{47} &\textcolor{cyan}{144.2778} &\textcolor{cyan}{32.7236} & 0.0231 $\pm$ 0.0045 & 0.0313 $\pm$ 0.0060 & 0 $\pm$ 0.0053 & 0.7442 $\pm$ 0.0395 & 9.7106 $\pm$ 0.5049 & 14.3700 $\pm$ 0.7789 & 5.2882 $\pm$ 0.1123 \\
\textcolor{cyan}{49} &\textcolor{cyan}{144.2435} &\textcolor{cyan}{32.7246} & 0.0119 $\pm$ 0.0831 & 0.0426 $\pm$ 0.0484 & 0 $\pm$ 0.0268 & 0.2938 $\pm$ 0.0202 & 1.8468 $\pm$ 0.1020 & 2.6500 $\pm$ 0.3166 & 1.7170 $\pm$ 0.0706 \\
\textcolor{green}{50} &\textcolor{green}{144.2854} &\textcolor{green}{32.6877} & 0.0039 $\pm$ 0.0026 & 0.0221 $\pm$ 0.0039 & 0 $\pm$ 0.0039 & 0.0306 $\pm$ 0.0075 & 3.2687 $\pm$ 0.3817 & 2.9900 $\pm$ 0.4238 & 3.0087 $\pm$ 0.4635 \\
        \hline
    \end{tabular}
  \end{adjustbox}  
    \label{tab:IR_intensity}
\end{table*}
 
\begin{table*}
      \centering
\caption{Additional insights to the selected locations.}
\label{tab:my_label}
      \begin{threeparttable}
      \begin{adjustbox}{width=\textwidth, center}          
\begin{tabular}{ccccc}
        \hline
        Location No. & Nearby HII regions & Star forming complex\tnote{a} & Distance to the closest HII region (arcsec)\tnote{b} & Age of HII regions (Myr)\tnote{c}  \\
        \hline
         & Peak intensity at 100 $\micron$ & & \\
         \hline
         2 & HSK 10, 16, 20 & NW & 3.74 (HSK 20) &  3.5 - 6.3 \\
         3 & HSK 4, 6, 7 &  & 1.80 (HSK 7) & 2.5 - 3.5, 4.5 - 6.3 \\
         4 & HSK 61, 65, 67 & SE & 2.52 (HSK 65) & 2.5 - 4.5 \\
         7 & HSK 26 & NW & 3.96 (HSK 7) & 3.5 - 4.5 \\
         21 & HSK 71, 73 & NE & 4.15 (HSK 73) & 2.5 - 3.5 \\
         24 & HSK 50, 52 & N & 5.77 (HSK 50) & 3.5 - 4.5  \\
         27 & HSK 13 & NW & 6.13 (HSK 13) & 2.5 - 3.5  \\
         44 & HSK 15, 17 & NW & 2.17 (HSK 15) & 3.5 - 6.3 \\
         30 & HSK 63, 64, 70 & SE & 0.60 (HSK 70) & 3.5 - 4.5 \\
         47 & HSK 73, 74 & NE & 2.88 (HSK 74) & 2.5 - 3.5 \\
         49 & HSK 80, 81, 82 & Ext NE & 2.52 (HSK 80) & 3.5 - 4.5 \\
         \hline
         & Peak intensity at 70 $\micron$ & & \\
         \hline
         5 & HSK 45 & N & 17.28 (HSK 45) & 3.7\tnote{d} \\
         6 & HSK 35 & N & 12.60 (HSK 35) & 4.5 - 6.3 \\
         41 & HSK 57, 58 & SE & 1.87 (HSK 58) & 3.5 - 4.5 \\
         42 & HSK 7 & & 17.31 (HSK 7) & 2.5 - 3.5 \\
         46 & HSK 71 & NE & 2.24 (HSK 71) & 3.5\tnote{d} \\
         50 & HSK 31 & NW & 12.24 (HSK 31) & 6.3\tnote{d} \\
         \hline
         & Voids (N(HI) < $1 \times 10^{21}$ cm$^{-2}$) & & \\
         \hline
         1 & HSK 25, 31, 32 & NW & 9.11 (HSK 32) & 4.5 - 6.3 \\
         11 & HSK 31, 32, 35 & NW & 13.83 (HSK 31) & 4.5 - 6.3 \\
         19 & HSK 30 & NW & 0.73 (HSK 30) & 3.5 - 4.5 \\
         32 & HSK 10, 12 & NW & 20.16 (HSK 12) & 3.5 - 4.5 \\
         33 & HSK 47 & N & 13.38 (HSK 47) & 3.5 - 4.5 \\
         35 & HSK 35 & NW & 32.04 (HSK 35) & 4.5 - 6.3 \\
         36 & HSK 3, 5, 11 & Int. Shell & 26.28 (HSK 3) & 2.5 - 4.5 \\
         \hline
         & Peak intensity at 100 $\micron$ with 8 $\micron$ emission & & \\
         \hline
         8 & HSK 25 & NW & 5.76 (HSK 25) & 6.2\tnote{d} \\
         12 & HSK 39, 41, 45 & N & 4.43 (HSK 45) &  2.5 - 3.5 \\
         37 & HSK 49, 51 & N & 1.45 (HSK 49) & 4.5 - 6.3 \\
         \hline
    \end{tabular}
    \end{adjustbox}
    \begin{tablenotes}
        \item[a] The star forming complexes have been adopted from \citet{Egorov2017}.
        \item[b] The nearest HII region is indicated inside the parentheses.
        \item[c] Estimated from \citet{Stewart2000}.
        \item[d] Taken from \citet{Wiebe2014}.
    \end{tablenotes}
    \end{threeparttable}
\end{table*}

%%%%%%%%%%%%%%%%%%%%%%%%%%%%%%%%%%%%%%%%%%%%%%%%%%

% Don't change these lines
%\bsp	% typesetting comment
\label{lastpage}
\end{document}